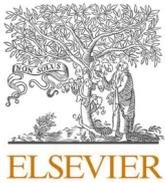
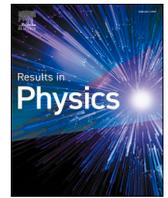

# Analysis of bio-nanofluid flow over a stretching sheet with slip boundaries

Bahram Jalili [a], Salar Ghadiri Alamdari [a], Payam Jalili [a],*, D.D. Gani [b],*

[a] *Department of Mechanical Engineering, Faculty of Engineering, North Tehran Branch, Islamic Azad University, Tehran, Iran*
[b] *Department of Mechanical Engineering, Babol Noshirvani University of Technology, P.O. Box 484, Babol, Iran*

ABSTRACT

A viscous, incompressible, micropolar bio-nanofluid flowing across a stretching sheet in three dimensions while being driven to convect several slip boundaries in the presence of a magnetic field was studied. With the assistance of the relevant transformations, a mathematical model is presented. The finite difference method numerically solves the converted non-linear ordinary differential equations. A comprehensive assessment was conducted to examine the impact of governing parameters on dimensionless velocity, micro-rotation, temperature, nanoparticle volume fraction, microorganisms, and heat transfer rate. The findings of this investigation showed a strong correlation when compared to previous studies, indicating a high level of agreement and consistency between the results. The study's conclusions indicate that the velocity profile increases with higher values of $\lambda$ and $\delta$, while it decreases with higher values of $A$, $\delta_v$, and $M$. The micro-rotation profile $F(\eta)$ drops as the spin gradient viscosity parameter rises, but $G(\eta)$ increases. The temperature profile decreases with higher Prandtl numbers but increases with higher thermophoresis parameters. The concentration profile decreases with higher Schmidt numbers and Brownian motion parameters. The microorganism profile increases with higher Peclet numbers and microorganism slip parameters but decreases with higher bio-convection Schmidt numbers. Lastly, the local Nusselt number grows with increasing values of the stretching parameter $\lambda$.

## Introduction

The fundamental component of all significant machine-made applications is the transfer of heat. When compared to the capacity of thermal systems and devices, the working fluids' thermal conductivity determines how much heat can be transferred. In traditional heat transfer, nanoparticles between 1 and 100 nm in size are liquefied to transport heat in fluids. Nitride ceramics, metals, and semiconductors make up the structure of nanofluid particles. It has been discovered that nanofluid makes a good thermal conductivity medium. The inclusion of nanoparticles enhances the base fluids' efficiency and thermal conductivity.

The utilization of ultra-fine solid particles in fluids has improved heat transfer significantly during the past few years, and one such technology is nanofluid. Choi and J.A. [1] coined the phrase "nanofluid." In 2007, Das et al. [2] described a variety of uses for nanofluids. The usage of nanofluid in a convective boundary layer flow was examined by Buongiorno [3]. Nanofluid behavior was explained using Brownian diffusion and thermophoresis. We looked at the Buongiorno model and discovered that the slip and base fluid velocities add up to the flow velocity. The four scales of a nanofluid—molecular, micro, macro, and megascale—were explored by Wang and Fan [4]. Zhang et al.'s study of the MHD radiative flow of a nanofluid over a surface with varying heat flux and chemical reaction is found in [5]. Muhammad et al. [6] investigated the processes involved in the ferromagnetic fluid's heat transfer across a stretching sheet with thermal stratification. Makinde and Aziz [7] reported the study of nanofluid flowing over a stretching sheet under the convective boundary condition with boundary layer flow. They investigated the impact of convective heating at the sheet on higher temperatures as well as greater rates of heat transfer from the sheet. Nadeem and Lee examined the nanofluid [8] while boundary layer flow was occurring across a stretching sheet.

Fluids with microstructures are known as micropolar fluids. The non-symmetric stress tensor is to which they belong. Micropolar fluid is made up of stiff, spherical, or randomly oriented particles. They spin and micro-rotate independently while suspended in a viscous fluid. Polar fluid includes micropolar fluid. They can rotate, shrink, and have other microsize effects. Blood flow, bubbly liquids, liquid crystals, and other phenomena are physical examples of micropolar flow. Xu and Pop [9] have examined a recent discovery of micropolar fluid that combines nanofluid with bioconvection development. Aziz et al. [10] conducted a theoretical analysis of the natural bio-convection flow through the

**Nomenclature**

| | |
|---|---|
| $a$ | stretching–shrinking rate based on surface velocity |
| $A$ | unsteadiness parameter |
| $A_1^*$ | first Rivlin-Ericksen tensor |
| $b^*$ | stretching–shrinking rate based on surface temperature |
| $b_1$ | stretching–shrinking rate based on surface concentration |
| $b_2$ | stretching–shrinking rate based on surface microorganism |
| $\tilde{b}$ | chemotaxis constant |
| $B$ | micro-inertia density parameter |
| $C_{fx}$ | local skin friction coefficient |
| $C$ | nanoparticle volume fraction |
| $C_w$ | wall nanoparticle volume fraction |
| $C_\infty$ | ambient nanoparticle volume fraction |
| $D_B$ | Brownian diffusion coefficient |
| $D_T$ | thermophoretic diffusion coefficient |
| $D_m$ | microorganism diffusion coefficient |
| $D_1(x,y)$ | local thermal slip factor |
| $(D_1)_0$ | constant thermal slip factor |
| $E_1(x,y)$ | local microorganisms slip factor |
| $(E_1)_0$ | constant microorganisms slip factor |
| $f(\eta)$ | dimensionless stream function (along the x-axis) |
| $F(\eta)$ | dimensionless stream function (along the y-axis) |
| $j$ | micro-inertia density |
| $n$ | number of motile microorganism |
| $n_w$ | wall motile microorganisms |
| $n_\infty$ | ambient motile microorganism |
| $N$ | component of the micro-rotation vector normal to the × - y plane |
| $N_1(x,y)$ | local velocity slip factor |
| $(N_1)_0$ | constant velocity slip factor |
| $Nb$ | Brownian motion parameter |
| $Nt$ | thermophoresis parameter |
| $Nu_x$ | local Nusselt number |
| $Pe$ | bioconvection Peclet number |
| $Pr$ | Prandtl number |
| $q_n$ | surface microorganism flux |
| $q_w$ | surface heat flux |
| $Re_x$ | local Reynolds number |
| $Sb$ | bioconvection Schmidt number |
| $Sc$ | Schmidt number |
| $t$ | dimensional time |
| $T$ | nanofluid temperature |
| $T_w$ | wall temperature |
| $T_\infty$ | ambient temperature |
| $u$ | velocity component along the x-axis |
| $u_e$ | external fluid velocity |
| $u_w$ | wall velocity |
| $v$ | velocity component along the y-axis |
| $W_c$ | maximum cell swimming speed |
| $\alpha$ | effective thermal diffusivity |
| $\alpha_0$ | dimensional constant |
| $\delta_v$ | velocity slip parameter |
| $\delta_T$ | thermal slip parameter |
| $\delta_n$ | microorganism slip parameter |
| $\delta$ | micropolar parameter |
| $\eta$ | independent similarity variable |
| $\theta(\eta)$ | dimensionless temperature |
| $\kappa$ | vortex viscosity coefficient |
| $\lambda$ | stretching–shrinking constant |
| $\lambda_0$ | spin-gradient viscosity parameter |
| $\mu$ | dynamic viscosity of the fluid |
| $\nu$ | kinematic viscosity |
| $\rho$ | nanofluid density |
| $\sigma$ | spin-gradient viscosity |
| $\tau$ | ratio of the effective heat capacity of the nanoparticle material to the fluid heat capacity |
| $\tau_w$ | shear stress |
| $\varphi(\eta)$ | dimensionless nanoparticle volume fraction |
| $\chi(\eta)$ | dimensionless number of motile microorganism |
| $\psi$ | stream function |

*Subscripts*

| | |
|---|---|
| $()'$ | ordinary differentiation with respect to $\eta$ |

boundary layer of a nanofluid. In order to explore a micropolar fluid over a stretching sheet, Agarwal et al. [11] used finite element flow and heat transfer simulations. Hassanien and Gorla [12] studied the micropolar steady flow through the boundary layer of porous and impermeable sheets. The theory of micropolar fluids was put forth by Eringen [13]. El-Aziz [14] looked at whether or not micropolar fluids with individual motion could maintain rotation. Spin inertia's effects help to support bodily moments and stress.

A collective phenomenon is the bio-convection arrangement. They are primarily caused by the upswimming of slightly thick microorganisms, which are incredibly minute living things like bacteria that may be seen under a microscope. The up-swimming causes the top layer of the fluid layers of suspensions to become overly dense. It became unstable because bacteria gathered there. The microbes descend, resulting in bioconvection. When it comes to geophysical phenomena, such as the phenomenon of thermophiles, where mobile microorganisms migrate to hot springs, thermo-bio-convection is crucial. Another goal is to use microbes to boost oil recovery in the field. Oil-bearing strata are given nutrients and microorganisms to establish porosity deviation. In light of this, Nield and Kuznetsov [15] claimed that microorganisms might aid in micro-system growth, as they exhibit a substantial aspect in mixing and a rise in mass mobility. The occurrence of Stefan blowing and the impact of multiple slips on the bio-convection flow of a nanofluid across a moving plate was examined by Uddin et al. [16]. Uddin et al. [17] conducted a study on the flow of nanofluids through a flat vertical plate under convective boundary layer conditions without considering Magnetohydrodynamics (MHD) effects. The researchers focused on a Newtonian heating boundary condition. Khan et al. [18] investigated how gyrotactic microorganisms flowed across the boundary layer as they passed a vertical plate featuring a magnetic field. According to Kuznetsov [19], thermo-bio-convection started to occur in a shallow fluid-saturated porous layer that was heated from below in an oxytactic microorganism solution. Jalili et al. [20] investigated the flow characteristics of a micropolar ferrofluid over a contracting surface with enhanced thermal conductivity, focusing on the influence of magnetic and boundary factors. The outcomes demonstrated that the magnetic and boundary parameters influence velocity similarly in contrast to the micro-rotation parameter. Jalili et al. [21] employed a unique combination of analytical and numerical methods to investigate the heat transfer behavior within a cylindrical polar system subjected to a magnetic field. They have suggested novel approaches to problem-solving.

Hamid et al. [22] explored the combined effects of fluctuating magnetic field and heat generation/absorption on unsteady flow using a stretching cylinder to produce a non-Newtonian Williamson fluid in the presence of nanoparticles. For the most recent revision of Buongiorno's nanofluid model, they considered the effects of binary chemical reactions and activation energy. Hamid et al. [23] looked into the Williamson fluid flow produced by stretching and contracting a sheet in addition to ohmic heating. Compared to the homogeneous model, the obtained results indicated a higher agreement of this model with experimental data. Yang et al.'s [24] goal was to investigate the impact of a non-uniform heat source/sink on the Carreau fluid flow via a permeable stretching/shrinking sheet at the unsteady stagnation point. Additional effects of magnetohydrodynamics, joule heating, and viscous dissipation were added to the flow model to increase its originality. The thermal radiation aspect of the bioconvection flow of magnetized Sisko nanofluid along a stretching cylinder with swimming microorganisms was investigated by Yin et al. [25]. The research showed that the thermophoresis diffusion parameter reduced the thermal boundary layer thickness and temperature field. In their study, Zhang et al. [26] investigated the numerical simulation of the bioconvection radiative flow of Williamson nanofluid across a vertical stretching cylinder with activation energy and swimming microorganisms. The Williamson nanofluid bioconvection flow past a stretching cylinder embedded in the Darcy-Forchheimer medium was studied for its rheological properties. The Darcy-Forchheimer flow of the Casson bioconvection fluid in the presence of the chemical reaction influence was theoretically investigated by Rehman et al. [27]. Investigations were conducted into the effects of Soret and Dufour on electroosmotic forces (EOF).

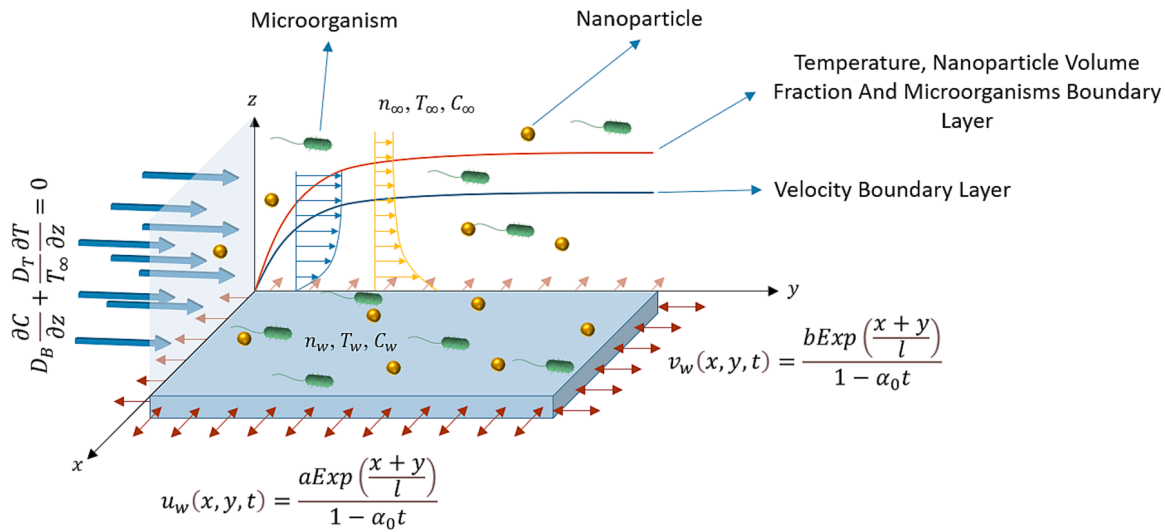

**Fig 1.** Geometry of the problem.

We explored the parameters of mass and heat transmission in our analysis. We have investigated a micropolar nanofluid flow that is unstable, laminar, and incompressible in three dimensions. The partial differential equations were converted through similarity transformation into non-linear ordinary differential equations. The Finite Difference Method is used to solve these equations numerically. Different parameter influences are explored graphically. Additionally, the graphical behavior of the Nusselt number, the microorganism flux, and the behavior of skin friction, as investigated numerically, are discussed.

**Governing equations**

The analysis of a microorganism-containing nanofluid flowing across an exponentially stretched sheet is investigated. Fig. 1 depicts the problem's geometry. The velocity field components in the ×, y, and z directions are denoted as u, v, and w, respectively. Within the boundary layer, the symbols n, T, and C represent the density of motile microorganisms, temperature, and nanoparticle volume fraction, respectively. In contrast, the motile microorganism density, fluid temperature, and nanoparticle volume friction are represented by the letters $n_w$, $T_w$, and $C_w$, respectively, at the wall. $n_\infty$, $T_\infty$, and $C_\infty$, on the other hand, are used to denote these quantities away from the wall.

Multiple slip and zero mass flux states exist on the plate's surface. By taking into account the aforementioned supposition, the necessary equations for motion, energy, nanoparticles, and motile microorganisms are defined as [3,15],

$$\nabla \bullet V = 0 \qquad (1)$$

$$\frac{\rho dV}{dt} = \mathrm{div}\tau + J \times B \qquad (2)$$

$$\tau = -pI + \mu A_1^* \qquad (3)$$

*I* **Stands for the identity tensor in equation (3), p stands for pressure, μ for dynamic viscosity, and $A_1^*$ is used to represent the first Rivlin-Ericksen tensor. In mathematics, $A_1^*$ is written as**

$$A_1^* = \nabla \bullet V + (\nabla \bullet V)^T \qquad (4)$$

As a result, following simplification, we obtain the necessary momentum equation as a partial differential equation and micropolar as,

$$\rho \frac{DV}{Dt} = -\nabla p + (\mu + \kappa)\nabla^2 V + \kappa(\nabla \times N) \qquad (5)$$

$$\rho j \frac{DV}{Dt} = \gamma \nabla^2 N - \kappa(2N - \nabla \times V) \qquad (6)$$

In this issue, the energy equation is,

$$\frac{DT}{Dt} = \alpha \nabla^2 T + (\tau)[D_B \nabla T \bullet \nabla C + (\frac{D_T}{T_\infty})\nabla T \bullet \nabla T] \qquad (7)$$

The concentration equation is,

$$\frac{DC}{Dt} = D_B \nabla^2 C + (\frac{D_T}{T_\infty})\nabla^2 T \qquad (8)$$

And the microorganism equation is,

$$\frac{Dn}{Dt} + \frac{\widetilde{b}W_c}{\nabla C}[\nabla n \bullet \nabla C] = D_m(\nabla^2 n) \qquad (9)$$

Here, $\frac{D}{Dt} = \frac{\partial}{\partial t} + u\frac{\partial}{\partial x} + v\frac{\partial}{\partial y} + w\frac{\partial}{\partial z}$ is the convective derivative about time, and ρ is the density of the nanofluid. N denotes the micro-rotation vector, V denotes the velocity vector, α the diffusivity constant, and γ the spin gradient viscosity. τ is the ratio of a nanoparticle's heat capacity to a fluid, $\widetilde{b}$ the chemotaxis constant, and $W_c$ the maximal cell speed. Micro-rotation viscosity coefficient is κ, micro-inertia density is j. The Brownian, thermophoretic, and microorganisms' diffusion coefficients are denoted by the letters $D_B$, $D_T$, and $D_m$, respectively.

In this model, the fluid velocity, temperature, and nanoparticle volume percentage vary with respect to the x, y, and t coordinates at the wall., which are displayed below.

$$u_w(x,y,t) = \frac{aExp\left(\frac{x+y}{l}\right)}{1-\alpha_0 t}, v_w(x,y,t) = \frac{bExp\left(\frac{x+y}{l}\right)}{1-\alpha_0 t}, T_w = T_\infty + \frac{b^*Exp\left(\frac{x+y}{2l}\right)}{(1-\alpha_0 t)^2}, C_w$$
$$= C_\infty + \frac{b_1 Exp\left(\frac{x+y}{2l}\right)}{(1-\alpha_0 t)^2}, and\, n_w = n_\infty + \frac{b_2 Exp\left(\frac{x+y}{2l}\right)}{(1-\alpha_0 t)^2}$$
(10)

Here, the constants $\alpha_0$; a and b all have a dimension per unit of time. The constants for microorganisms, the volume fraction of nanoparticles, and temperature are $b_2$, $b_1$, and $b^*$, respectively.

Currently, the flow problem's velocity field is,

$$V = [u(x,y,z,t), v(x,y,z,t), w(x,y,z,t)] \quad (11)$$

The component form of the equations above are as follows, utilizing these presumptions and the approximate boundary layer: $O(u) = 1$, $O(v) = 1, O(w) = \delta, O(x) = 1, O(y) = 1, O(z) = \delta$.

$$\frac{\partial u}{\partial x} + \frac{\partial v}{\partial y} + \frac{\partial w}{\partial z} = 0 \quad (12)$$

$$\frac{\partial u}{\partial t} + u\frac{\partial u}{\partial x} + v\frac{\partial u}{\partial y} + w\frac{\partial u}{\partial z} = \left(\frac{\mu+\kappa}{\rho}\right)\frac{\partial^2 v}{\partial z^2} - \frac{\kappa}{\rho}\frac{\partial N_2}{\partial y} - \frac{\sigma}{\rho}B^{*2}u \quad (13)$$

$$\frac{\partial v}{\partial t} + u\frac{\partial v}{\partial x} + v\frac{\partial v}{\partial y} + w\frac{\partial v}{\partial z} = \left(\frac{\mu+\kappa}{\rho}\right)\frac{\partial^2 u}{\partial z^2} - \frac{\kappa}{\rho}\frac{\partial N_1}{\partial y} - \frac{\sigma}{\rho}B^{*2}v \quad (14)$$

$$\rho j\left(\frac{\partial N_1}{\partial t} + u\frac{\partial N_1}{\partial x} + v\frac{\partial N_1}{\partial y} + w\frac{\partial N_1}{\partial z}\right) = \gamma\frac{\partial^2 N_1}{\partial z^2} - \kappa\left(2N_1 + \frac{\partial v}{\partial z}\right) \quad (15)$$

$$\rho j\left(\frac{\partial N_2}{\partial t} + u\frac{\partial N_2}{\partial x} + v\frac{\partial N_2}{\partial y} + w\frac{\partial N_2}{\partial z}\right) = \gamma\frac{\partial^2 N_2}{\partial z^2} - \kappa\left(2N_2 - \frac{\partial u}{\partial z}\right) \quad (16)$$

$$\frac{\partial T}{\partial t} + u\frac{\partial T}{\partial x} + v\frac{\partial T}{\partial y} + w\frac{\partial T}{\partial z} = \alpha\frac{\partial^2 T}{\partial z^2} + \tau D_B\frac{\partial T}{\partial z}\frac{\partial C}{\partial z} + \tau\frac{D_T}{T_\infty}\left(\frac{\partial T}{\partial z}\right)^2 \quad (17)$$

$$\frac{\partial C}{\partial t} + u\frac{\partial C}{\partial x} + v\frac{\partial C}{\partial y} + w\frac{\partial C}{\partial z} = D_B\frac{\partial^2 C}{\partial z^2} + \frac{D_T}{T_\infty}\frac{\partial^2 T}{\partial z^2} \quad (18)$$

$$\frac{\partial n}{\partial t} + u\frac{\partial n}{\partial x} + v\frac{\partial n}{\partial y} + w\frac{\partial n}{\partial z} = D_m\frac{\partial^2 n}{\partial z^2} - \frac{\widetilde{b}W_c}{C_w - C_\infty}\left[\frac{\partial}{\partial z}\left(n\frac{\partial C}{\partial z}\right)\right] \quad (19)$$

Relevant boundary conditions for the problem are expected to take the form of,

$$u = \lambda u_w(x,t) + \nu N_1^*(x,y,t)\left(\frac{\partial u}{\partial z}\right), v = \lambda v_w(y,t) + \nu N_2^*(x,y,t)\left(\frac{\partial v}{\partial z}\right), N_1 = n\frac{\partial v}{\partial z}, N_2$$
$$= -n\frac{\partial u}{\partial z}, T = T_w(x,y,t) + D_1(x,y,t)\left(\frac{\partial T}{\partial z}\right), D_B\frac{\partial C}{\partial z} + \frac{D_T}{T_\infty}\frac{\partial T}{\partial z} = 0, n$$
$$= n_w(x,y,t) + E_1(x,y,t)\left(\frac{\partial n}{\partial z}\right), \text{ when } z \to 0$$
(20)

$$u = 0, v = 0, N \to 0, T \to T_\infty, C \to C_\infty, n \to n_\infty, \text{ when } z \to \infty \quad (21)$$

In this part, we define the velocity, temperature, and microorganism slip factors, respectively, by, $N_1^*(x,y,t) = (N_1)_0\sqrt{\frac{l(1-\alpha_0 t)}{\nu a}}, N_2^*(x,y,t) = (N_2)_0\sqrt{\frac{l(1-\alpha_0 t)}{\nu a}}, D_1(x,y,t) = (D_1)_0\sqrt{\frac{l(1-\alpha_0 t)}{\nu a}}, and\, E_1(x,y,t) = (E_1)_0\sqrt{\frac{l(1-\alpha_0 t)}{\nu a}}$. Assuming mass flux to be zero, we used the formula $D_B\frac{\partial C}{\partial z} + \frac{D_T}{T_\infty}\frac{\partial T}{\partial z} = 0$. Specifically defined dimensionless variables of the following form are introduced as,

$$\eta = z\sqrt{\frac{a}{\nu l(1-\alpha_0 t)}}Exp\left(\frac{x+y}{2l}\right), N_1 = F(\eta)\sqrt{\frac{a^3}{\nu l(1-\alpha_0 t)^3}}Exp\left(\frac{3(x+y)}{2l}\right), N_2$$
$$= G(\eta)\sqrt{\frac{a^3}{\nu l(1-\alpha_0 t)^3}}Exp\left(\frac{3(x+y)}{2l}\right), T = T_\infty + \frac{b^*Exp\left(\frac{x+y}{2l}\right)}{(1-\alpha_0 t)^2}\theta(\eta), C$$
$$= C_\infty + \frac{b_1 Exp\left(\frac{x+y}{2l}\right)}{(1-\alpha_0 t)^2}\phi(\eta), n = n_\infty + \frac{b_2 Exp\left(\frac{x+y}{2l}\right)}{(1-\alpha_0 t)^2}\chi(\eta), \theta = \frac{T-T_\infty}{T_w-T_\infty}, \phi$$
$$= \frac{C-C_\infty}{C_w-C_\infty}, \chi = \frac{n-n_\infty}{n_w-n_\infty}$$
(22)

Here, $\eta$ is the similarity variable, and $f(\eta), g(\eta), F(\eta), G(\eta), \theta(\eta), \phi(\eta), \chi(\eta)$, which stand for linear velocity, microrotation, temperature, nanoparticle volume fraction, and microbes, correspondingly, are dimensionless variables. The following are the velocity's components:

$$u = \frac{af'(\eta)Exp\left(\frac{x+y}{l}\right)}{(1-\alpha_0 t)}, v = \frac{ag'(\eta)Exp\left(\frac{x+y}{l}\right)}{(1-\alpha_0 t)}, w$$
$$= -(f+\eta f' + g + \eta g')\sqrt{\frac{a\nu}{2l(1-\alpha_0 t)}}Exp\left(\frac{x+y}{2l}\right)$$
(23)

Prime indicates the differentiation with regard to g in the aforementioned Eq. (23), where the velocity elements in the ×, y, and z directions are denoted by u, v, and w, accordingly. By transforming the system of equations above using similarity variables,

$$(1+\delta)f''' + (f+g)f'' - 2(f'+g')f' - A(2f'+\eta f'') - Mf' - \delta G' = 0 \quad (24)$$

$$(1+\delta)g''' + (f+g)g'' - 2(f'+g')g' - A(2g'+\eta g'') - Mg' - \delta F' = 0 \quad (25)$$

$$\lambda_0 F'' + (f+g)F' - 3(f'+g')F - A(3F+\eta F') - \delta B(2g''+4F) = 0 \quad (26)$$

$$\lambda_0 G'' + (f+g)G' - 3(f'+g')G - A(3G+\eta G') + \delta B(2f''+4G) = 0 \quad (27)$$

$$\frac{1}{Pr}\theta'' + (f+g)\theta' - (f'+g')\theta - A(4\theta+\eta\theta') + Nb\theta'\phi' + Nt\theta'^2 = 0 \quad (28)$$

$$\frac{1}{Sc}\phi'' + (f+g)\phi' - (f'+g')\phi - A(4\phi+\eta\phi') + \frac{Nt}{Nb}\frac{1}{Sc}\theta'' = 0 \quad (29)$$

$$\frac{1}{Sb}\chi'' + (f+g)\chi' - (f'+g')\chi - A(4\chi+\eta\chi') - \frac{Pe}{Sb}(\chi\phi'' + \chi'\phi') = 0 \quad (30)$$

The pertinent boundary conditions are stated as follows:

$$f'(\eta) = \lambda + \delta_v f''(\eta), f(\eta) = 0, g'(\eta) = \lambda c + \delta_v g''(\eta), g(\eta) = 0, F(\eta)$$
$$= ng''(\eta), G(\eta) = -nf''(\eta), \theta(\eta) = 1 + \delta_T\theta'(\eta), Nt\theta'(\eta) + Nb\phi'(\eta)$$
$$= 0, \chi(\eta) = 1 + \delta_n\chi'(\eta) \text{ when } \eta \to 0$$
(31)

$$f'(\eta) = 0, g'(\eta) = 0, F(\eta) = 0, G(\eta) = 0, \theta(\eta) = 0, \phi(\eta) = 0, \chi(\eta)$$
$$= 0 \text{ when } \eta \to \infty$$
(32)

The parameters in the preceding system of differential equations are denoted by $\delta = \frac{\kappa}{\mu}$ for the micropolar parameter, $A = \frac{l\alpha_0}{a}$ for unsteadiness, $c = \frac{b}{a}$ for stretching, $M = \frac{\sigma l B^{*2}}{\rho u_w}$ for the magnetic field, $B = \frac{\nu l}{j u_w}$ for micro-inertia density, and $\lambda_0 = \frac{\gamma}{j\nu}$ for spin gradient viscosity, n stands for the micro-gyration, $Nb = \frac{\tau D_B\Delta C}{\nu}$ for Brownian motion, $Nt = \frac{\tau D_T\Delta T}{\nu T_\infty}$ for thermophoresis, $Sc = \frac{\nu}{D_B}$ for Schmidt number, $Sb = \frac{\nu}{D_m}$ for bio-convection Schmidt number, $Pe = \frac{bw_C D_m}{\nu^2}$ for bio-convection Peclet number, and $Pr = \frac{\nu}{\alpha}$ for Prandtl number. The velocity slip, thermal slip, and microorganisms slip factor are also defined as,

$$\delta_v = (N_1^*)_0\nu, \delta_T = (D_1)_0, \delta_n = (E_1)_0 \quad (33)$$

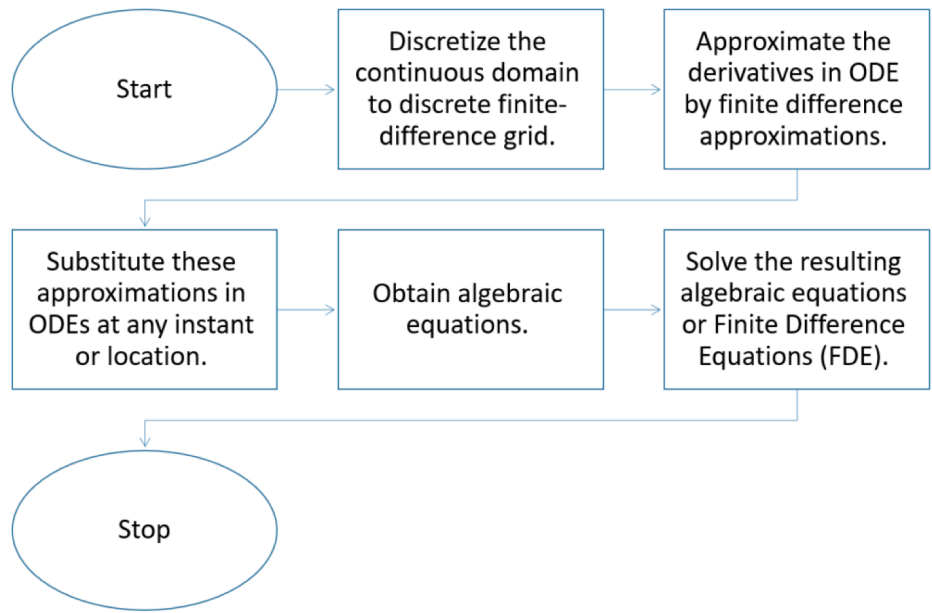

Fig 2. Flow chart of the finite difference method.

$N_1^*$, $D_1$, and $E_1$ are proportional to $\sqrt{\frac{l(1-\alpha_0 t)}{\nu a}}$, as can be seen. In the aforementioned equations, c and $\lambda$ are the constants of stretching, where $\lambda < 0$ denotes shrinkage, $\lambda = 0$ denotes stationary motion, and $\lambda > 0$ denotes plate stretching.

*Physical quantities*

Physical quantities play a crucial role in engineering as they provide valuable insights into various aspects such as flow characteristics, heat transfer rates, and microorganism transfer rates. These quantities enable engineers to analyze and understand the behavior and performance of systems, allowing for efficient design, optimization, and decision-making processes. By quantifying and studying these physical parameters, engineers can gain a comprehensive understanding of the underlying processes and phenomena, leading to advancements in various fields of engineering. Skin friction coefficient, local Nusselt number, and local motile microorganisms' flux are defined as, respectively, $C_{fx}$ and $C_{fy}$ in the directions of × and y.

$$C_{fx} = \frac{\tau_{wx}}{\rho u_w^2}, C_{fy} = \frac{\tau_{wy}}{\rho u_w^2}, Nu_x = \frac{xq_w}{T_w - T_\infty}, Q_{nx} = \frac{xj_w}{D_m n_w}, \tau_{wx}$$
$$= 2[(\mu+\kappa)\frac{\partial u}{\partial z}\Big|_{z=0} + \kappa(N_2)_{z=0}], \tau_{wy} = 2[(\mu+\kappa)\frac{\partial v}{\partial z}\Big|_{z=0} + \kappa(N_1)_{z=0}], q_w$$
$$= -k\frac{\partial T}{\partial z}\Big|_{z=0} \quad vj_w = -D_B\frac{\partial n}{\partial z}\Big|_{z=0}$$
(34)

The shear stresses in the directions of × and y are, respectively, $\tau_{wx}$ and $\tau_{wy}$ in the equations above. Thermal conductivity is k. The flux of both surface heat and mobile microorganisms is $q_w$ and $j_w$, respectively. Applying the transformation of the similarity variables mentioned above, we obtain,

$$Re_x^{1/2} C_{fx} = 2(1+\delta(1-n))f''(0) \quad (35)$$

$$Re_x^{1/2} C_{fy} = 2(1+\delta(1+n))g''(0) \quad (36)$$

$$Re_x^{-1/2} Nu_x = -\theta'(0) \quad (37)$$

$$Re_x^{-1/2} Q_{nx} = -\chi'(0) \quad (38)$$

The Local Reynolds number is indicated by $Re_x = u_w\sqrt{\frac{l(1-\alpha_0 t)}{\nu a}}$.

**Methodology**

The finite difference method has been utilized to solve Eq. (24) to Eq. (30) with Eq. (31) and Eq. (32) as boundary conditions. The numerical solutions of differential equations based on finite differences provide the values at discrete grid points. The idea behind FDM is to use algebraic difference techniques instead of the derivatives included in the governing differential equations. This generates an algebraic equation system that can be solved using accepted analytical or numerical techniques to determine the dependent variables' values at the discrete grid points.

The current value at $\eta = \eta_i$ and the forward step at $\eta_{i+1} = \eta_i + \Delta\eta_i$ are used in this approach to express the rate of change of the function about the variable $\eta$. The derivative of the function $f(\eta)$ can be represented

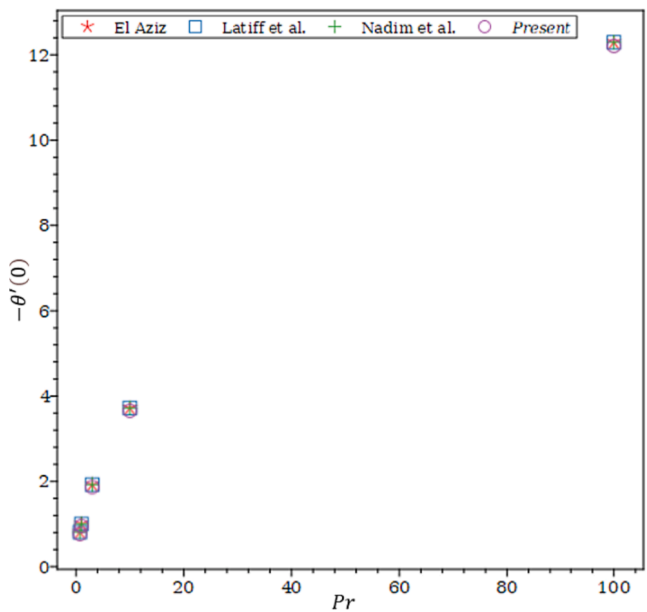

Fig 3. $-\theta'(0)$ values for various Pr.

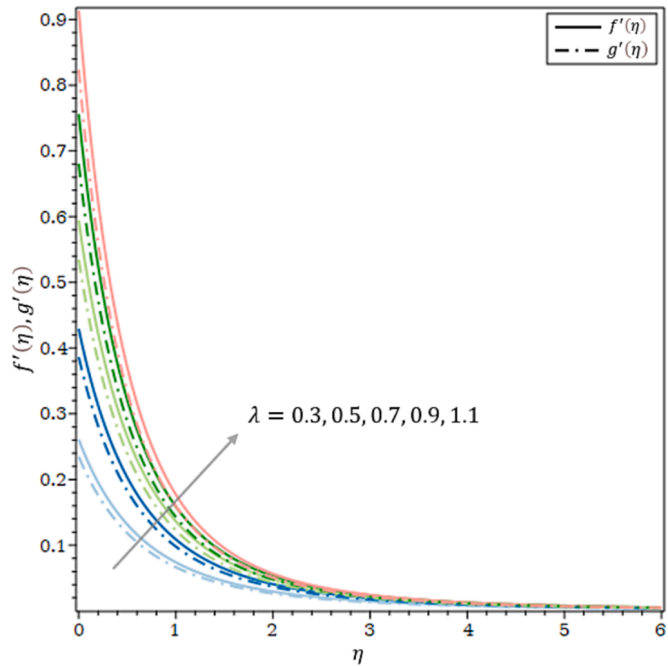

**Fig 4.** Velocity profiles for varied $\lambda$ values.

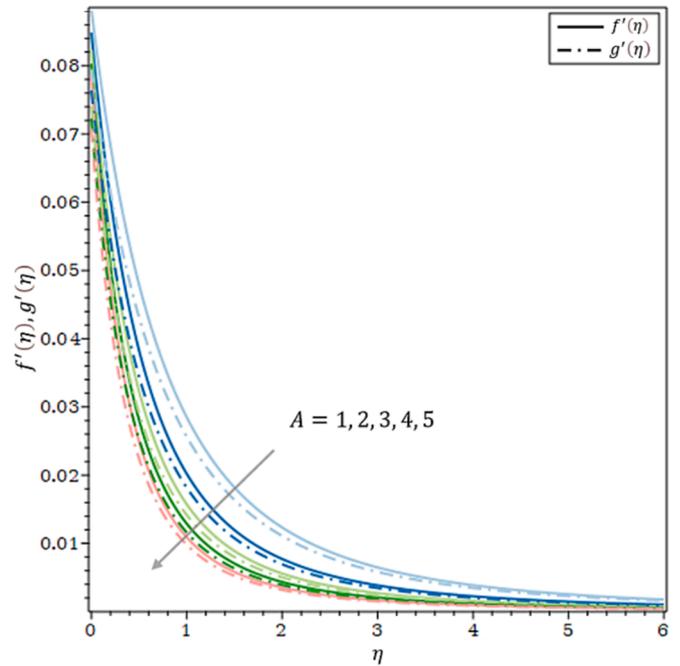

**Fig 5.** Velocity profiles for varied $A$ values.

mathematically as

$$\Delta f_i = \frac{df(\eta)}{d\eta}\bigg|_{\eta=\eta_i} \approx \frac{f_{i+1}-f_i}{\eta_{i+1}-\eta_i} = \frac{f_{i+1}-f_i}{\Delta\eta} = \frac{f_{i+1}-f_i}{h}, \Delta f_{i+1} = \frac{f_{i+2}-f_{i+1}}{h}, \Delta f_{i+2}$$
$$= \frac{f_{i+3}-f_{i+2}}{h}, etc.$$

(39)

where h represents the step size increase.

It is possible to obtain the second-order derivative of the function at $\eta$ as

$$\Delta^2 f_i = \frac{d}{d\eta}\left(\frac{df(\eta)}{d\eta}\right)\bigg|_{\eta=\eta_i} = \frac{\Delta f_{i+1} - \Delta f_i}{h} = \frac{f_{i+2}-2f_{i+1}+f_i}{h^2} \quad (40)$$

The same process can be used to derive higher-order derivatives.

Fig. 2 is the flow chart outlining the steps involved in solving ordinary differential equations (ODEs) using a finite-difference method, a numerical approach to approximate solutions for ODEs. This flow chart illustrates the process of solving ODEs, which is particularly useful when analytical solutions are difficult or impossible. By discretizing the domain, approximating derivatives, and solving resulting algebraic equations, you can obtain numerical solutions to ODEs for various scientific and engineering applications.

Fig. 3 illustrates a comparison between our research findings and previous studies regarding the values of $-\theta'(0)$ for different Prandtl numbers. It should be noted that this comparison is conducted under the assumption of disregarding the concentration and microorganism equation and the velocity and thermal slip boundary conditions. By focusing solely on the $-\theta(0)$ values and considering variations in the Prandtl number, our results provide insights into how they align or differ from previous research findings. Results from FDM and earlier research had a good degree of consistency.

**Result and discussion**

The flow behavior can be comprehensively assessed by employing a mathematical model and conducting computational simulations. This study conducted a parametric analysis to explore various aspects of the flow. The numerical results encompass dimensionless quantities such as velocity, microrotation, temperature, volume fraction of nanoparticles, microorganisms, and heat transfer rate. These results offer valuable insights into the characteristics and trends of the flow behavior, enabling a deeper understanding of the system under investigation.

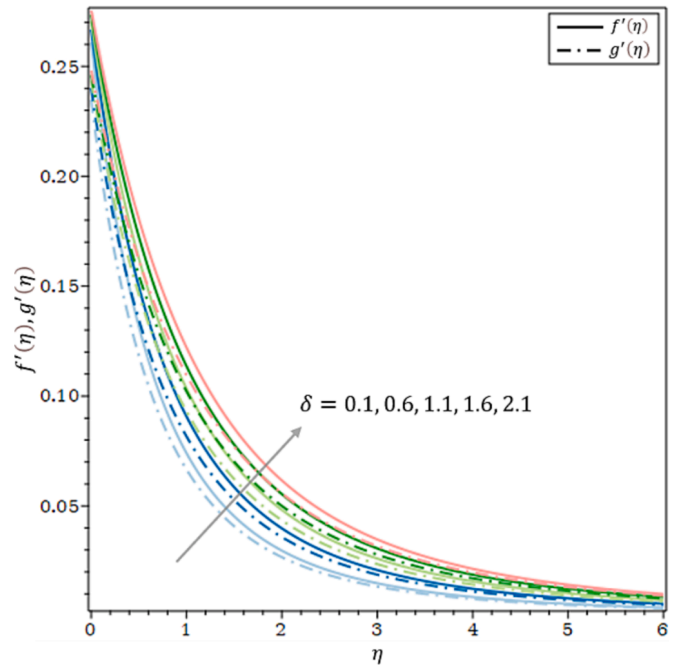

**Fig 6.** Velocity profiles for varied $\delta$ values.

The stretching's effects on the dimensionless velocity profile are illustrated in Fig. 4. The visualization reveals that the boundary layer's accelerated flow is caused by the momentum effect of a stretched sheet, which adds additional forces to the fluid's velocity. As $\lambda$ rises, velocity profiles in both directions increase.

As depicted in Fig. 5, the velocity field decreases as the value of the unsteadiness parameter $A$ increases. This indicates an inverse correlation between the unsteadiness parameter and the stretching velocity rate. Specifically, the unsteadiness parameter tends to decrease as the stretching velocity rate increases. These observations highlight the interrelationship between these two parameters and provide insights into the system's dynamics under study.

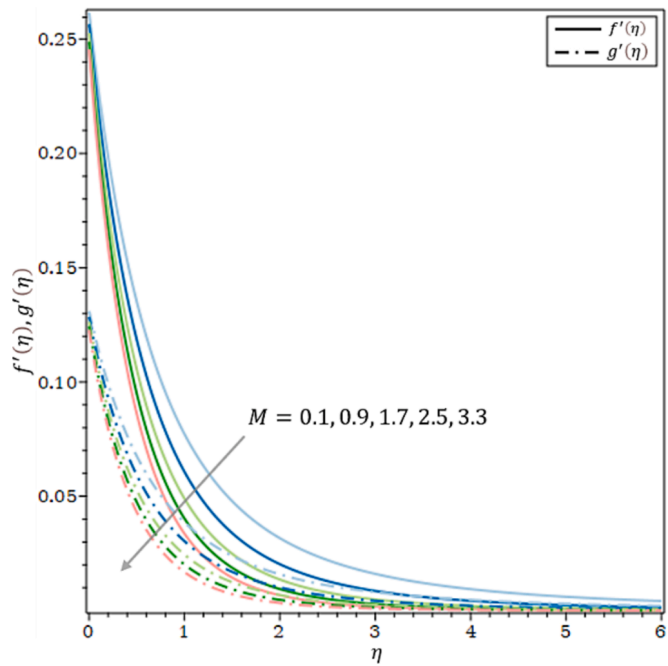

**Fig 7.** Velocity profiles for varied $M$ values.

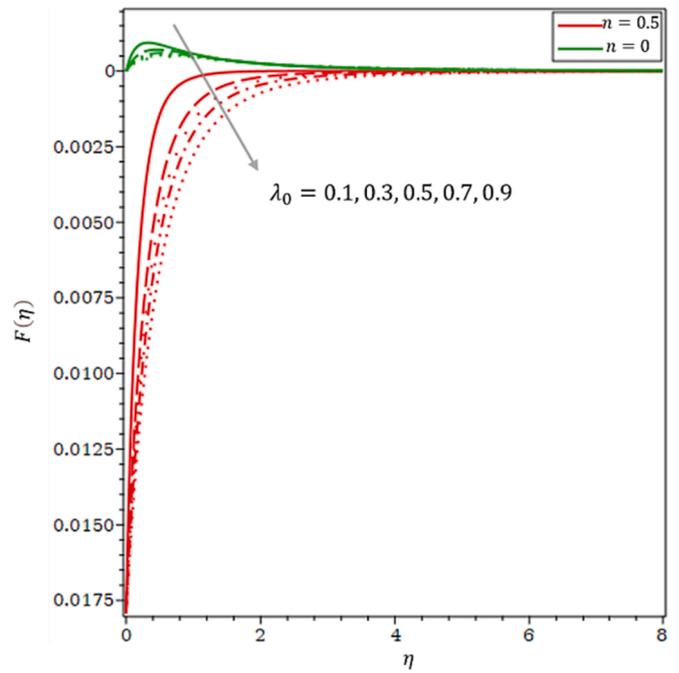

**Fig 9.** Microrotation profiles for varied $\lambda_0$ values.

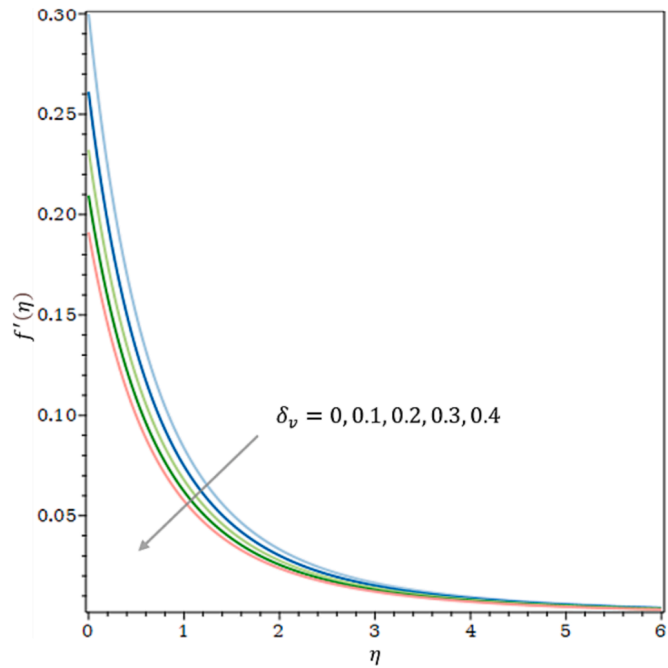

**Fig 8.** Velocity profiles (in the × direction) for varied $\delta_v$ values.

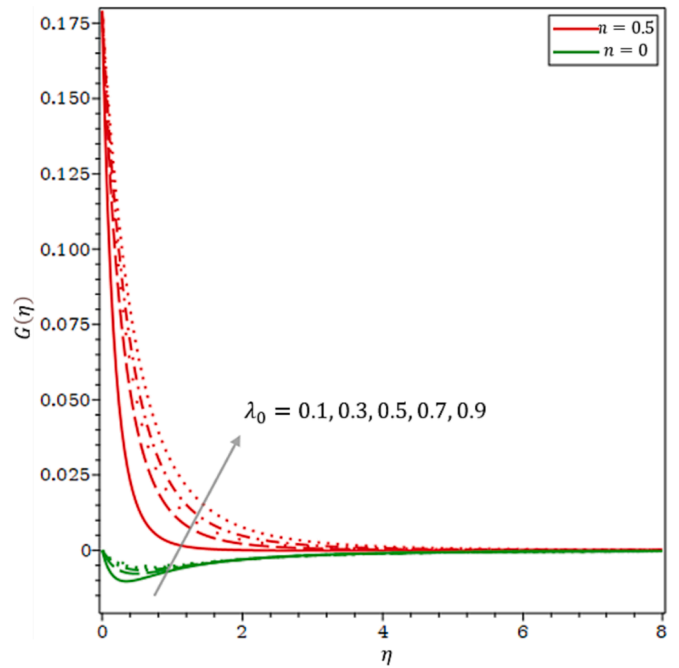

**Fig 10.** Microrotation profiles for varied $\lambda_0$ values.

Fig. 6 illustrates the influence of the micropolar parameter $\delta$ on the velocity field. The velocity distribution has been perceived to grow as the micropolar parameter does. The dynamic viscosity and the micropolar parameter are directly correlated; the micropolar parameter rises as the dynamic viscosity does.

The illustration in Fig. 7 is the impact of the magnetic field on velocity. Due to an increase in the magnetic field parameter, the velocity field decreases. The $c < 1$ is what causes the velocity profiles of the × and y directions to diverge from one another. The stretching value of velocities in both directions gets more similar as the c rises from 0 to 1.

Fig. 8 illustrates that the slip parameter $\delta_v$ increases, the velocity profile (in the × direction) decreases. As is evident, they are inversely related to one another.

In Figs. 9 and 10, the influence of the spin gradient viscosity parameter $\lambda_0$ on the micro-rotation profile is illustrated. When $\lambda_0$ is raised, the micro-rotation profile in Fig. 9 declines, but in Fig. 10, the micro-rotation profile rises. Here, n (micro-gyration parameter) is the difference in particle concentration, where $n = 0$ implies dense particle concentration, where the microelements are incapable of spinning near the wall, $n = \frac{1}{2}$ stands for poor particle concentration, and $n = 1$ refers to the flow of turbulent boundary layers.

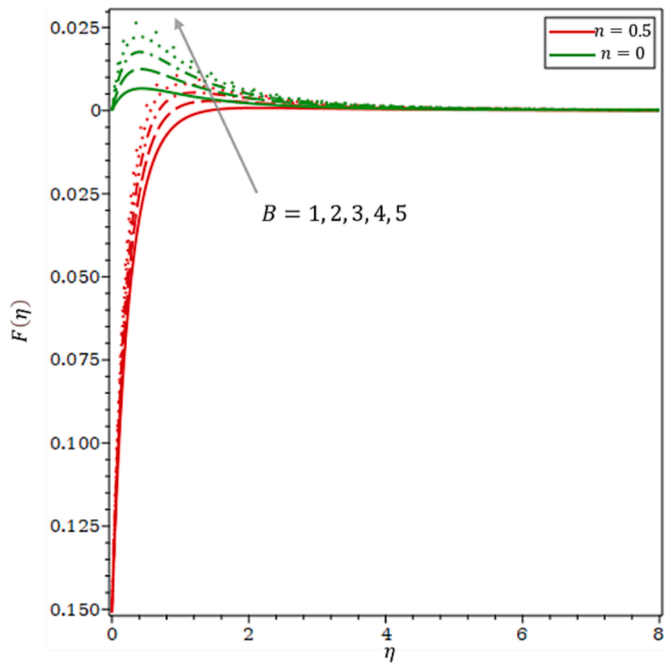

**Fig 11.** Microrotation profiles for varied *B* values.

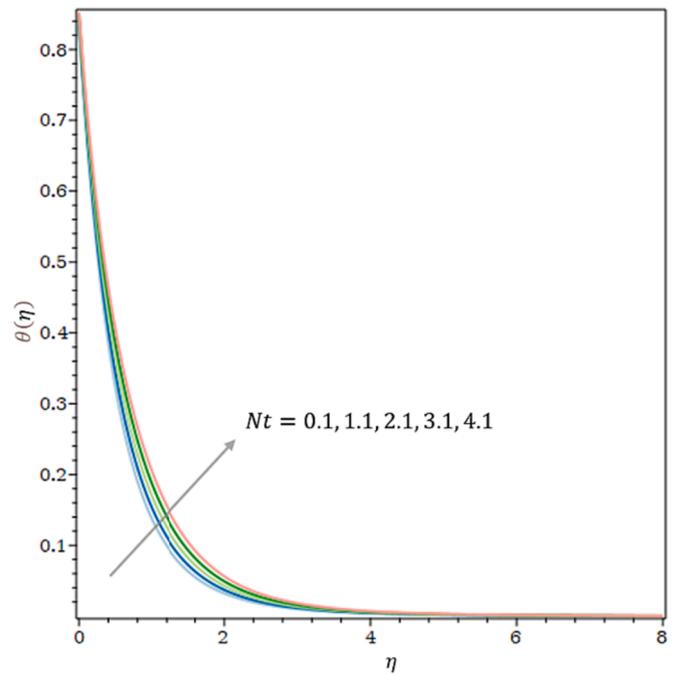

**Fig 13.** Temperature profiles for varied *Nt* values.

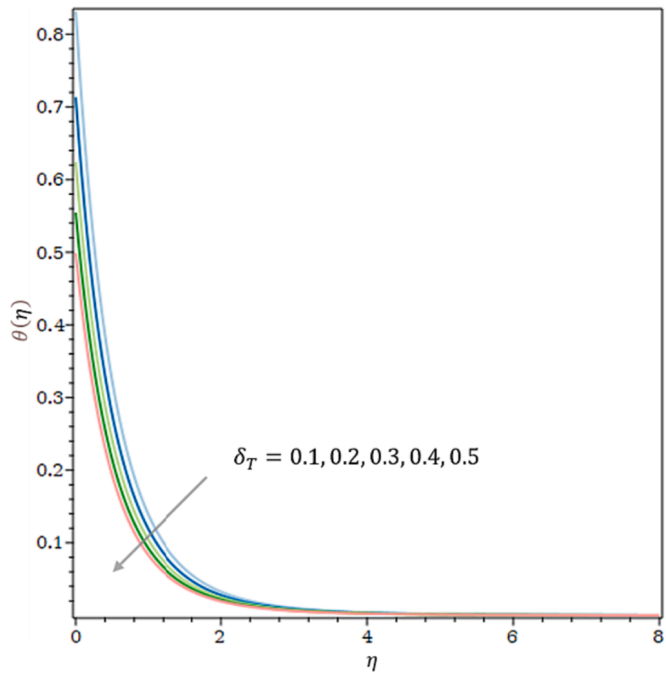

**Fig 12.** Temperature profiles for varied $\delta_T$ values.

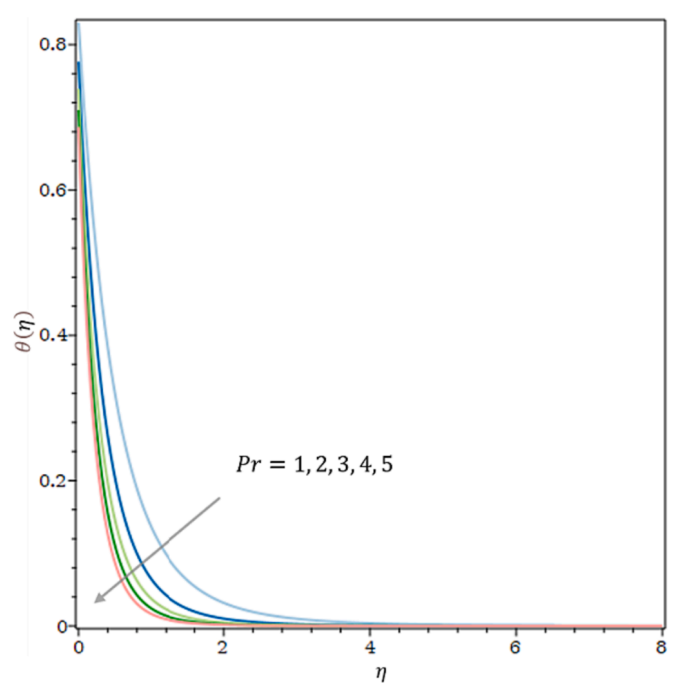

**Fig. 14.** Temperature profiles for varied *Pr* values.

Fig. 9 shows that the microrotation profile reduces when n moves from 0 to 0.5, or from dense to weak particle concentration. However, there is a direct relationship between them in Fig. 10.

In Fig. 11, the microrotation profile is displayed, illustrating the impact of the microinertia density parameter *B*. As *B* increases, the microrotation profile exhibits enhancement, indicating an improvement in the distribution and magnitude of microrotation within the system. Additionally, the microrotation profile decreases when the parameter n increases from 0 to 0.5. This observation suggests that an increase in n leads to a reduction in the microrotation effects.

Fig. 12 depicts the temperature field's change for various values of the thermal slip parameter $\delta_T$. With an increase in the thermal slip parameter, a drop in the temperature field is observed.

The influence of the thermophoresis parameter *Nt* on temperature is seen in Fig. 13. It demonstrates how rising temperatures and thicker thermal boundary layers are caused by raising the value of *Nt*.

Fig. 14 illustrates how the Prandtl number *Pr* affects the temperature profile. It demonstrated that an increase in the value of *Pr* causes a decrease in temperature and boundary layer thickness. It indicates that the decrease in thermal diffusivity is caused by the bigger value of *Pr*.

The influence of the Brownian motion parameter *Nb* on concentra-tion is illustrated in Fig. 15. The increase in Brownian parameter *Nb* results in a decrease in the concentration profile.

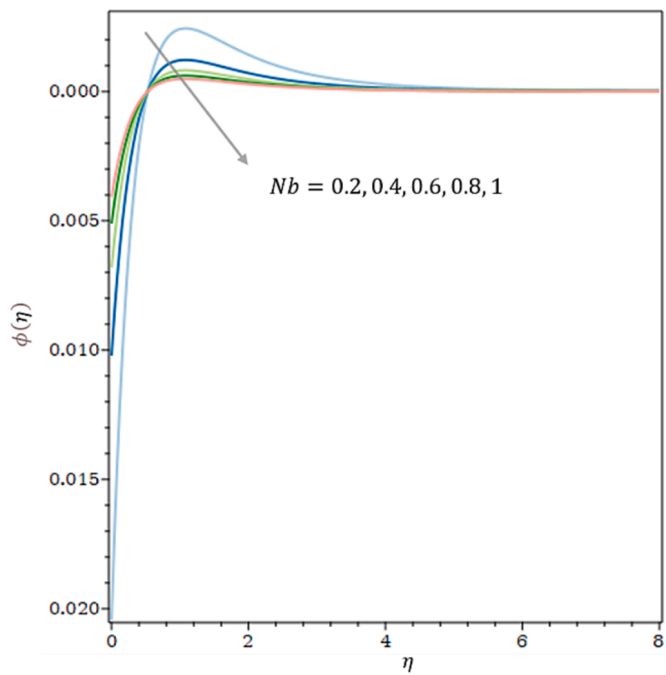

**Fig 15.** Nanoparticle volume fraction profiles for varied *Nb* values.

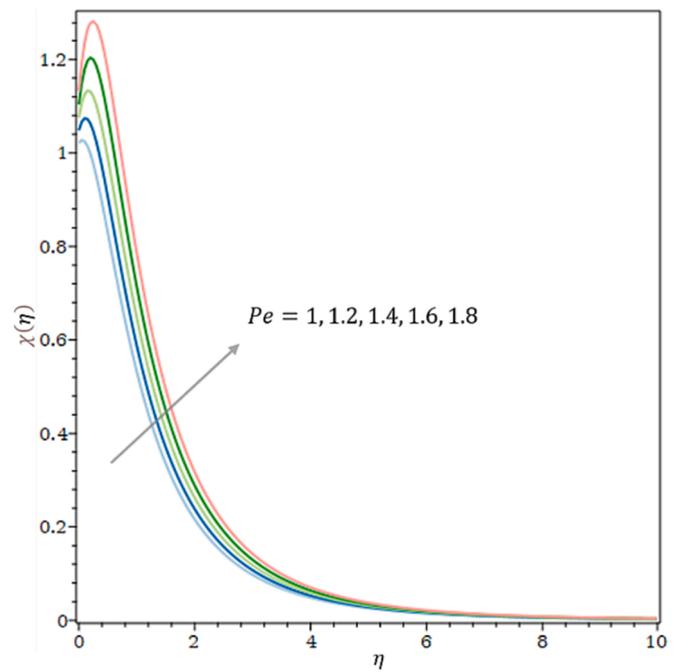

**Fig 17.** Microorganism profiles for varied *Pe* values.

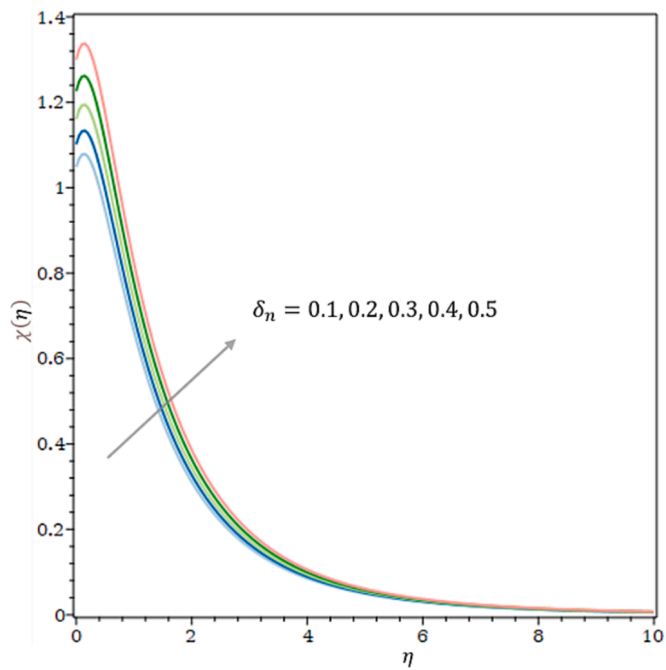

**Fig 16.** Microorganism profiles for varied $\delta_n$ values.

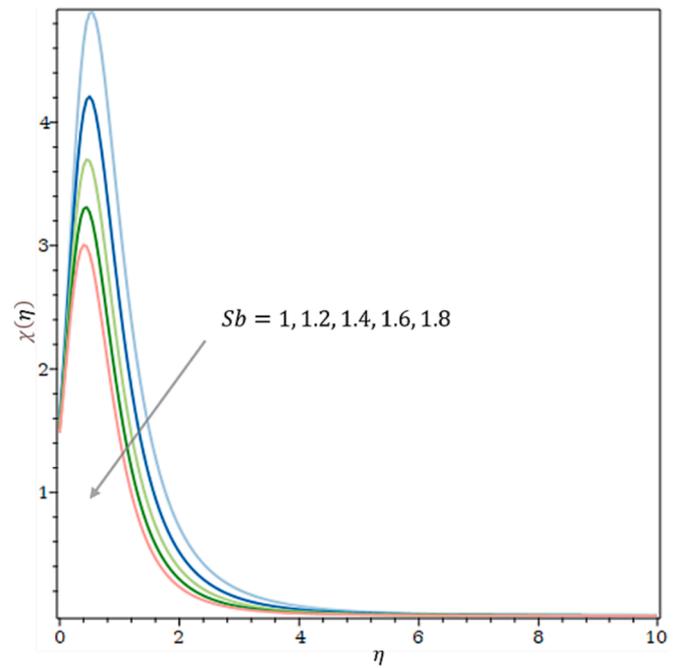

**Fig 18.** Microorganism profiles for varied *Sb* values.

Fig. 16 reports the influences of the microorganism slip parameter $\delta_n$ on microorganisms. When the microorganism slip factor increases, the density of motile microorganisms increases.

Figs. 17 and 18 depict the influence of the Peclet number (*Pe*) and the bio-convection Schmidt number (*Sb*) on the microorganism profile. The observations reveal that as the Peclet number, which relates to the diffusivity of microorganisms and their swimming motion, increases, the microorganism profile also increases. This indicates a direct relationship between *Pe* and the microorganism profile. On the other hand, when the bio-convection Schmidt number (*Sb*) increases, the microorganism profile decreases, while the motile microorganism density and the boundary layer thickness also decrease. This demonstrates an inverse relationship between *Sb* and the microorganism profile. These findings highlight the significant impact of *Pe* and *Sb* on the behavior of microorganisms within the system, shedding light on their diffusivity, swimming motion, and chemotaxis.

Fig. 19 shows how the stretching parameter affects the Nusselt number. Fig. 19 demonstrates how the heat transfer rate increased when the stretching parameter increased. The stretched sheet transfers heat at a faster pace than the shrinking sheet. The negative sign in the context indicates that heat is transferred from regions of higher temperature to regions of lower temperature when the sheet undergoes shrinkage or stretching. It is noted that these parameters do not impact the heat transfer rate of the velocity slip and microorganisms slip parameters. The variations in the heat transfer rate are independent of changes in the velocity slip and microorganism slip parameters.

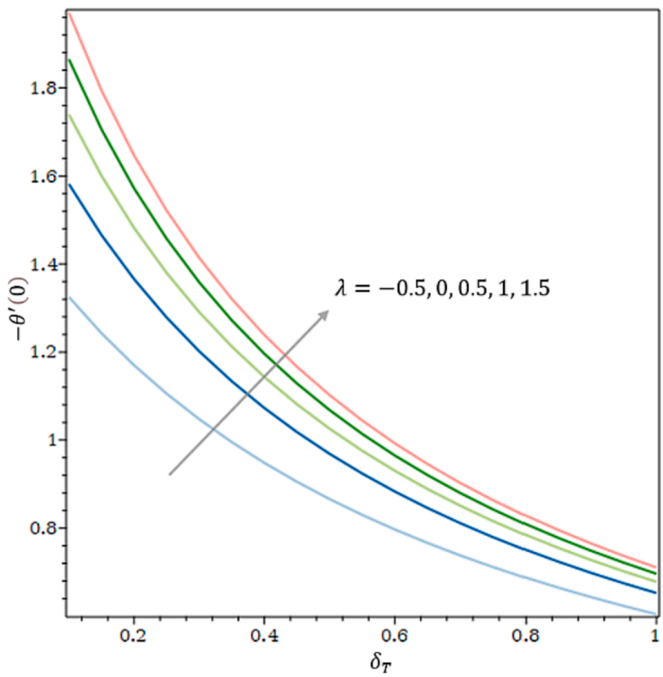

**Fig 19.** Heat transfer rate profiles for varied $\lambda$ values.

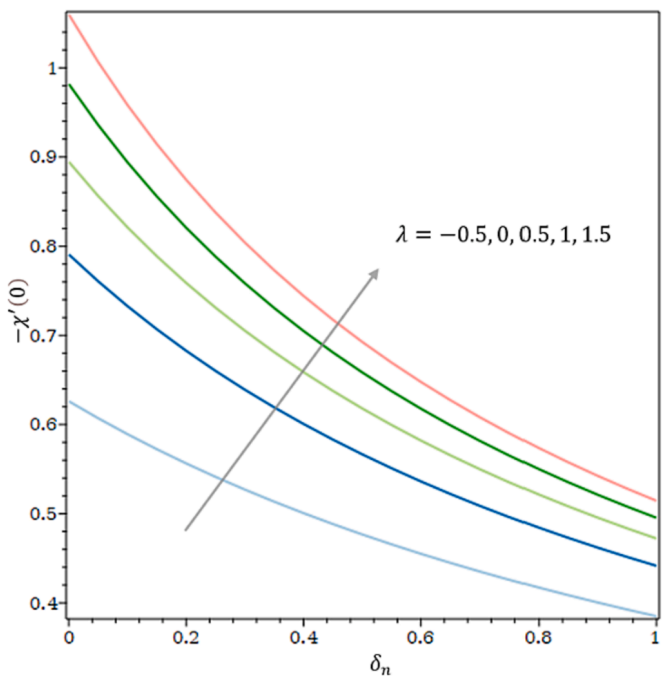

**Fig 20.** Microorganism transfer rate profiles for varied $\lambda$ values.

The microorganism transfer rate's relationship to the stretching parameter is depicted in Fig. 20. The microorganism transfer rate increased as the stretching parameter rose, as seen in Fig. 20. Microorganisms are transferred more quickly from the stretched sheet than from the decreasing sheet. The context's negative sign shows that when the sheet shrinks or stretches, microorganisms are transported from areas with a higher density of to those with a lower density. It should be observed that the microorganism transfer rate is unaffected by the velocity slip and thermal slip parameters. Modifying the velocity slip and temperature slip parameters did not affect the microorganism transfer rate.

Table 1 compares the skin friction coefficients in the × and y axes for varying magnetic field values, micropolar, unsteadiness, and micro-inertia density parameters in the × and y axes. The parameters given above range from 0.1 to 0.7.

**Table 1**
Comparison of $Re_x^{1/2} C_{fx}$ and $Re_x^{1/2} C_{fy}$ for varied values of $M$, $\delta$, $A$ and $B$.

| M | δ | A | B | $Re_x^{1/2} C_{fx}$ | $Re_x^{1/2} C_{fy}$ |
|---|---|---|---|---|---|
| 0.1 | 0.1 | 0.1 | 0.2 | 1.21978045664804 | 1.22104866613984 |
| 0.2 | | | | 1.25001060403651 | 1.25131184063634 |
| 0.3 | | | | 1.27922426082879 | 1.28055702135731 |
| | 0.2 | | | 1.34331368642763 | 1.34509169663944 |
| | 0.3 | | | 1.40342739477690 | 1.40639238627263 |
| | 0.4 | | | 1.42683673150855 | 1.47176428543347 |
| | | 0.2 | | 1.50990505433540 | 1.51390787097660 |
| | | 0.3 | | 1.55807212002422 | 1.56142649802767 |
| | | 0.4 | | 1.60438640382987 | 1.60738274394023 |
| | | | 0.1 | 1.60785273277246 | 1.60980088366903 |
| | | | 0.3 | 1.60036604479890 | 1.60521421877089 |
| | | | 0.5 | 1.58975523881717 | 1.60159480090867 |
| | | | 0.7 | 1.57175242096679 | 1.59926431335209 |

**Conclusion**

A 3D unsteady forced convection multiple slip boundary layer flow of a viscous incompressible micropolar bio-nanofluid over a stretching sheet in the presence of a magnetic field was studied. The proper transformation was used to convert a mathematical model into a collection of related ordinary differential equations. The finite difference approach was used to resolve these equations. The results obtained from the study demonstrate the influence of various factors on the velocity, temperature, micro-rotation, nanoparticle volume fractions, and motile microorganisms within the system. Additionally, a graph depicting the heat transfer rate is provided. Based on these findings, the key aspects of the flow problem can be summarized as follows:

- The velocity profile increases as the values of the parameters $\lambda$ and $\delta$ increase, while it decreases as the values of the parameters $A$, $\delta_v$, and $M$ change.
- The micro-rotation profile $F(\eta)$ drops while profile $G(\eta)$ increases as the value of the spin gradient viscosity parameter rises. The micro-rotation profile $F(\eta)$ also increases by incrementing the value of $B$ for the micro-inertia parameter.
- When the values of the Prandtl number and the thermophoresis parameter increase, the temperature profile exhibits a decrease for the Prandtl number and an increase for the thermophoresis parameter.
- The concentration profile decreases as the Schmidt number and Brownian motion parameter increase in value.
- The microorganism profile increases as the Peclet number and the microorganism slip parameter increase. Conversely, it decreases when the bio-convection Schmidt number increases.
- The local Nusselt number and microorganism transfer rate grows as the stretching parameter's value, $\lambda$, is increased.

For upcoming attempts, the authors advise doing analyses of various geometries as well as utilizing analytical solutions to the equations.

**Declaration of Competing Interest**

The authors declare that they have no known competing financial interests or personal relationships that could have appeared to influence the work reported in this paper.

**Data availability**

Data will be made available on request.

**References**

[1] Choi, S. U. S., & Eastman, J. (1995). Enhancing thermal conductivity of fluids with nanoparticles (Vol. 66).
[2] Das SK, Choi SUS, Yu W, Pradeep T. Nanofluids: science and technology. Willey, Hoboken 2007. https://doi.org/10.1002/9780470180693.
[3] Buongiorno J. Convective transport in nanofluids. J Heat Transfer 2005;128(3): 240–50. https://doi.org/10.1115/1.2150834.
[4] Wang L, Fan J. Nanofluids research: key issues. Nanoscale Res Lett 2010;5(8):1241. https://doi.org/10.1007/s11671-010-9638-6.
[5] Zhang C, Zheng L, Zhang X, Chen G. MHD flow and radiation heat transfer of nanofluids in porous media with variable surface heat flux and chemical reaction. App Math Model 2015;39(1):165–81. https://doi.org/10.1016/j.apm.2014.05.023.
[6] Muhammad N, Nadeem S, Haq RU. Heat transport phenomenon in the ferromagnetic fluid over a stretching sheet with thermal stratification. Results Phys 2017;7:854–61. https://doi.org/10.1016/j.rinp.2016.12.027.
[7] Makinde OD, Aziz A. Boundary layer flow of a nanofluid past a stretching sheet with a convective boundary condition. Int J Therm Sci 2011;50(7):1326–32. https://doi.org/10.1016/j.ijthermalsci.2011.02.019.
[8] Nadeem S, Lee C. Boundary layer flow of nanofluid over an exponentially stretching surface. Nanoscale Res Lett 2012;7(1):94. https://doi.org/10.1186/1556-276X-7-94.
[9] Xu H, Pop I. Mixed convection flow of a nanofluid over a stretching surface with uniform free stream in the presence of both nanoparticles and gyrotactic microorganisms. Int J Heat Mass Transf 2014;75:610–23. https://doi.org/10.1016/j.ijheatmasstransfer.2014.03.086.
[10] Aziz A, Khan WA, Pop I. Free convection boundary layer flow past a horizontal flat plate embedded in porous medium filled by nanofluid containing gyrotactic microorganisms. Int J Therm Sci 2012;56:48–57. https://doi.org/10.1016/j.ijthermalsci.2012.01.011.
[11] Agarwal RS, Bhargava R, Balaji AVS. Finite element solution of flow and heat transfer of a micropolar fluid over a stretching sheet. Int J Eng Sci 1989;27(11): 1421–8. https://doi.org/10.1016/0020-7225(89)90065-7.
[12] Hassanien IA, Gorla RSR. Heat transfer to a micropolar fluid from a non-isothermal stretching sheet with suction and blowing. Acta Mech 1990;84(1):191–9. https://doi.org/10.1007/BF01176097.
[13] Eringen AC. Theory of micropolar fluids. J Math. Mech. 1966;16:1–18. https://doi.org/10.1512/iumj.1967.16.16001.
[14] Abd El-Aziz M. Mixed convection flow of a micropolar fluid from an unsteady stretching surface with viscous dissipation. J Egyptian Math Soc 2013;21(3): 385–94. https://doi.org/10.1016/j.joems.2013.02.010.
[15] Nield DA, Kuznetsov AV. The cheng-minkowycz problem for the double-diffusive natural convective boundary layer flow in a porous medium saturated by a nanofluid. Int J Heat Mass Transf 2011;54(1):374–8. https://doi.org/10.1016/j.ijheatmasstransfer.2010.09.034.
[16] Uddin MJ, Kabir MN, Bég OA. Computational investigation of Stefan blowing and multiple-slip effects on buoyancy-driven bioconvection nanofluid flow with microorganisms. Int J Heat Mass Transf 2016;95:116–30. https://doi.org/10.1016/j.ijheatmasstransfer.2015.11.015.
[17] Uddin MJ, Khan WA, Ismail AI. MHD free convective boundary layer flow of a nanofluid past a flat vertical plate with newtonian heating boundary condition. PLoS One 2012;7(11):e49499. https://doi.org/10.1371/journal.pone.0049499.
[18] Khan WA, Makinde OD, Khan ZH. MHD boundary layer flow of a nanofluid containing gyrotactic microorganisms past a vertical plate with Navier slip. Int J Heat Mass Transf 2014;74:285–91. https://doi.org/10.1016/j.ijheatmasstransfer.2014.03.026.
[19] Kuznetsov AV. The onset of thermo-bioconvection in a shallow fluid saturated porous layer heated from below in a suspension of oxytactic microorganisms. Eur J Mech B Fluids 2006;25(2):223–33. https://doi.org/10.1016/j.euromechflu.2005.06.003.
[20] Jalili B, Jalili P, Sadighi S, Ganji DD. Effect of magnetic and boundary parameters on flow characteristics analysis of micropolar ferrofluid through the shrinking sheet with effective thermal conductivity. Chin J Phys 2021;71:136–50. https://doi.org/10.1016/j.cjph.2020.02.034.
[21] Jalili P, Ahmadi Azar A, Jalili B, Asadi Z, Domiri Ganji D. Heat transfer analysis in cylindrical polar system with magnetic field: a novel Hybrid Analytical and Numerical Technique. Case Studies Thermal Eng 2022;40:102524. https://doi.org/10.1016/j.csite.2022.102524.
[22] Hamid A, Hashim, Khan M. Impacts of binary chemical reaction with activation energy on unsteady flow of magneto-Williamson nanofluid. J Mol Liq 2018;262: 435–42. https://doi.org/10.1016/j.molliq.2018.04.095.
[23] Hamid A, Hashim K, Khan M, Hafeez A. Unsteady stagnation-point flow of Williamson fluid generated by stretching/shrinking sheet with Ohmic heating. Int J Heat Mass Transf 2018;126:933–40. https://doi.org/10.1016/j.ijheatmasstransfer.2018.05.076.
[24] Yang D, Israr Ur Rehman M, Hamid A, Ullah S. Multiple solutions for stagnation-point flow of unsteady carreau fluid along a permeable stretching/shrinking sheet with non-uniform heat generation. Coatings 2021;11(9):1012. https://doi.org/10.3390/coatings11091012.
[25] Yin J, Zhang X, Ur. Rehman I, Hamid A. Thermal radiation aspect of bioconvection flow of magnetized Sisko nanofluid along a stretching cylinder with swimming microorganisms. Case Studies Thermal Eng 2022;30:101771. https://doi.org/10.1016/j.csite.2022.101771.
[26] Zhang X, Yang D, Ur Rehman MI, Mousa AA, Hamid A. Numerical simulation of bioconvection radiative flow of Williamson nanofluid past a vertical stretching cylinder with activation energy and swimming microorganisms. Case Studies Thermal Eng 2022;33:101977. https://doi.org/10.1016/j.csite.2022.101977.
[27] Rehman MIU, Chen H, Hamid A, Guedri K, Abdeljawad T, Yang D. Theoretical investigation of darcy-forchheimer flow of bioconvection casson fluid in the presence of chemical reaction effect. Biomass Convers Biorefin 2022. https://doi.org/10.1007/s13399-022-03060-5.